# Superposition of Inductive and Capacitive Coupling in Superconducting LC Resonators

Sergiy Gladchenko, Moe Khalil, C. J. Lobb, F. C. Wellstood and Kevin D. Osborn

*Abstract* - We present an experimental investigation of lumped-element superconducting LC resonators designed to provide different types of coupling to a transmission line. We have designed four resonator geometries including dipole and quadrupole configured inductors connected in parallel with low loss $SiN_x$ dielectric parallel-plate capacitors. The design of the resonator allows a small change in the symmetry of the inductor or grounding of the capacitor to allow LC resonators with: 1) inductive coupling, 2) capacitive coupling, 3) both types of coupling, or 4) greatly reduced coupling. We measured all four designs at a temperature of 30mK at different values of power. We compare the extracted data from the four resonator types and find that both capacitive and inductive coupling can be included and that when left off, only a minor change in the circuit design is necessary. We also find a variation in the measured loss tangent of less than a few percent, which is a test of the systematic precision of the measurement technique.

*Index Terms* - Superconducting resonators; Dielectric losses; Electromagnetic coupling.

## I. INTRODUCTION

Superconducting resonators with parallel plate (lumped element) capacitors are present in a number of devices including phase qubits [1], amorphous dielectric resonators [2,3], and Josephson bifurcation amplifiers [4]. By design, they are compact and lack harmonic modes, and with a low-loss dielectric they can also achieve a high internal-quality factor at milli-Kelvin temperatures and low photon numbers [2,3].

The coupling between a qubit and a transmission lines must be made intentionally weak to achieve isolation from the environment. Variable coupling is sometimes used to achieve the necessary isolation/coupling between different qubits that are at the same resonant frequency [4,5,6]. At microwave frequencies, the nature of the coupling is not always intuitively obvious, and one often must resort to microwave modeling software. The coupling to the environment can include dc bias lines, as well as microwave input lines. In the case of a tunable coupler, it is very important that any (unintentional) fixed coupling is small so that the tunable coupler can achieve a high on-off ratio.

In resonators used to test dielectrics for application in quantum computing, it is especially important to understand the precision of the measurement technique. For the work described here the resonator is modeled as a LC circuit with finite loss and a coupling that can be capacitive, inductive coupling or both. While the same on-resonance coupling strength can be achieved in different ways, the nature of the coupling can affect the amount of noise that is coupled into the resonator off-resonance. In addition, if we include both inductive and capacitive components to the coupling, and vary their relative strength, the net magnitude of the coupling can change dramatically, so it is important to understand the inductive and capacitive components to the coupling in a design in order to achieve a desired overall coupling strength.

In this paper we report on measurements on resonators with different coupling types and different coupling values. The resonators all used a $SiN_x$ dielectric, which is known to show a power dependent loss at low temperatures. This will allow us to test of the accuracy of the theoretical circuit model and the precision of the measurement technique as a function of the ac amplitude across the capacitor.

## II. DEVICE DESCRIPTION AND FITTING MODEL

Figure 1. shows a schematic of a resonator with inductor L and capacitor $\hat{C}$ coupled to an input and output transmission line. The transmission lines carry waves traveling to the right with input amplitude $V^+_{in}$ and output amplitude $V^+_{out}$. Coupling to the LC resonator is achieved through a capacitor $C_c$ or a mutual inductance M. Four different resonator designs were tested with different types of coupling.

We first consider a resonator with capacitive coupling. In Fig.1 the capacitor $C_c$ represents the capacitive coupling between the center conductor of the resonator and one side of the capacitor. This capacitor allows energy to be coupled from the input transmission line to the LC resonator. On the other side of the LC resonator is a grounding wire (represented by a wire between two circular terminals in Fig. 1). As shown, this circuit shows capacitive coupling because current will be driven across the LC circuit on resonance. If however the grounding wire is not connected, no current can flow through the capacitor $C_c$ and the resonator is capacitively decoupled. In our real device, there will be stray capacitances (not shown) which allow some current to flow through $C_c$, but the device is nominally capacitively decoupled.

Manuscript received 3 August 2010. This research was performed at the Laboratory for Physical Sciences and C.J.Lobb acknowledges support from IARPA.

Sergiy Gladchenko and Kevin D. Osborn are with the Laboratory for Physical Sciences, College Park, MD 20740 USA. (e-mail: osborn@lps.umd.edu).

Moe S. Khalil is also with the University of Maryland, Department of Physics, College Park, MD 20740 USA.

C.J.Lobb and F. C. Wellstood are with the Joint Quantum Institute and the Center for Nanophysics and Advanced Materials, Department of Physics, University of Maryland, College Park, MD 20740 USA.

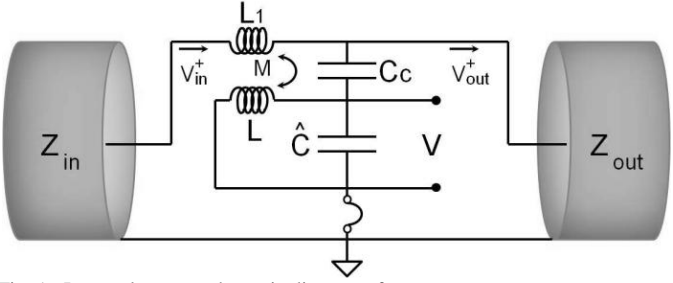

Fig. 1. Lump elements schematic diagram of resonator.

The other coupling type is inductive. In Fig. 1, $L_1$ couples magnetic flux into the inductor L, through mutual inductance M such that $M^2 \ll LL_1$. In our resonators we have two types of inductive coupling. In one case, the inductor is configured as a simple loop or dipole and the value of M is finite and appreciable. In the second case, L is configured as a first order gradiometer (quadrupole), such that the coupling M is nominally negligible.

A micrograph of a complete resonator with a quadrupole coil is shown in Fig. 2(a). All four resonator designs have the same capacitor layout, with a top plate size of 20μm by 325μm, which is intentionally small in one dimension to avoid trapping of magnetic vortices. The plates of the capacitor are connected to an inductor. A coplanar waveguide passes by the capacitor with a ground plane on either side of center conductor, which is labeled with arrows. The inductor and capacitor are embedded within a rectangular hole in the ground plane (not shown) of the CPW.

Fig.2(a) shows a resonator with the top plate of the capacitor connected to ground. This connection is made in the region in the upper dashed box, and Fig.2(b) shows a detailed view. For resonators where the capacitor is nominally decoupled, this connection is not present, as shown in Fig.2(c). The lower dashed box in Fig. 2(a) shows a quadropole inductor, and a separate view is shown in Fig. 2(d). This inductor has no nominal inductive coupling. In the inductively coupled designs this inductor is replaced with a simple meander loop or dipole inductor, as shown in Fig. 2(e).

The fabricated resonators realize all four possible combinations of these coupling types: capacitive coupling (C), inductive coupling (L), inductively and capacitive coupling (LC), and a nominally uncoupled resonator which has an imperfect (weak) value of residual coupling.

The devices were fabricated starting from an Al film that was sputtered on a sapphire substrate. We used optical lithography and a wet etch to pattern the base layer. An amorphous dielectric layer of $SiN_x$ was deposited on the base layer at 300°C using plasma-enhanced chemical vapor deposition (PECVD). Next a reactive ion etch (RIE) was used to open via through the dielectric. The upper wiring layer was then added by first using an ion mill to clean the exposed Al surface. A second Al layer was then sputtered and this was patterned a similar manner as the base layer.

### III. Measurement Technique

The resonators were mounted on a dilution refrigerator,

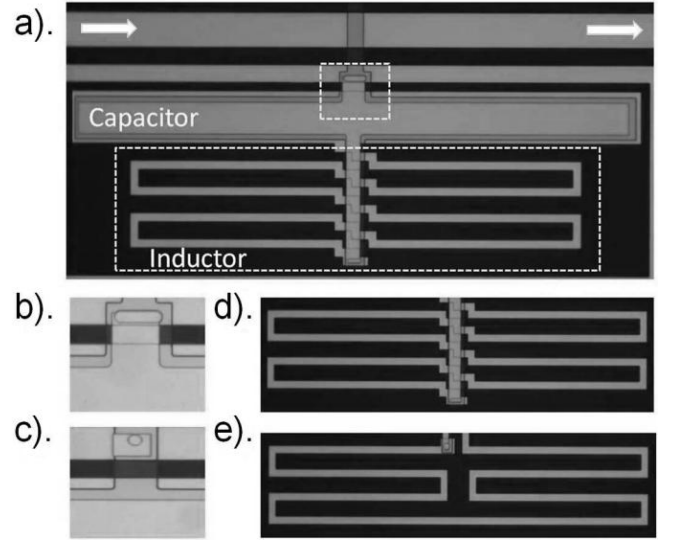

Fig. 2. Optical microscope images of resonator (light grey is metallization, black is sapphire substrate, dark grey is $SiN_x$). White squares in the (a) image indicate position of elements defining the type of coupling. Top square represents resonator grounding, bottom square show inductor. Images (b) - (e) demonstrated possible variation of these elements for different resonators. (b) - grounded resonator (top plate of capacitor connected to the grounded environment); (c) - non-grounded resonator; (d) - quadruple shaped loop of inductor; (e) - dipole shaped loop of inductor.

cooled to 30 mK, and the complex transmission amplitude $V^+_{out}/V^+_{in}$ was measured using a vector network analyzer. To reduce thermally generated noise from reaching the resonator, the input microwave line was attenuated by 20dB at both 1K and 30 mK. The output line has two Pamtech circulators at the same temperature stages to reduce the thermal photon background below a single photon within the bandwidth of the resonator. The transmitted signal is amplified with a HEMT amplifier at 4K and a low noise amplifier at room temperature.

The measured transmission amplitude versus frequency was fit to a model function derived from Fig. 1. For weak coupling we can make the approximations $L_1 \ll L \ll 1/(\omega_0^2 C_c)$ and we find

$$V^+_{out}/V^+_{in} = a\left(1 - \frac{Q_T/Q_e}{1 + i2Q_T(\omega - \omega_0)/\omega_0}\right) \quad (1),$$

where ω is applied microwave frequency, $\omega_0$ is the resonance frequency, $Q_T$ is the total quality factor, $Q_e$ is the external quality factor and $a$ is a constant nearly equal to unity.

To lowest order,

$$Re(1/Q_e) = \omega_0 L\left(\omega_0^2 C_c^2 Z_0 + \frac{M^2}{2Z_0 L^2}\right) \quad (2),$$

and the internal quality factor

$$Q_i \equiv Re(\hat{C})/Im(-\hat{C}) \quad (3),$$

is calculated from

$$1/Q_i = 1/Q_T - Re(1/Q_e) \quad (4),$$

Since here we will only attribute internal loss in the resonator to dielectric loss in the resonator capacitor, we can write

$$\hat{C} = C(1 - i\,tan\delta(V)), \quad (5)$$

where C is the real part of capacitance and $tan\delta$ is dielectric loss tangent. This gives an internal quality factor that is inversely proportional to the internal dielectric loss $Q_i^{-1} = tan\delta(V)$.

## IV. EXPERIMENTS AND DISCUSSION

Figure 3 represents the experimental result for $1/Q_e$ and $1/Q_i$. $1/Q_e$ describes losses due to the interaction of the resonator with the transmission line. Since this coupling is defined by the geometry of device, one expects this quantity to be independent of microwave power. This behavior is approximately observed for all four resonators.

Measured values for $Q_e$ and resonant frequency $f_r = \omega_0/2\pi$ for all four resonators are shown in Table I. As expected, the coupling is lowest for the nominally uncoupled resonator, highest for the LC coupled resonator, and in-between for the L and C coupled resonators. $Q_e$ differs by about a factor of 60 between the nominally uncoupled resonator and the LC coupled resonator. This shows that large changes in the coupling can be experimentally realized by making relatively small changes in the design. A more quantitative analysis is described below using microwave simulation software.

As discussed above, we model the internal loss in our resonators as dielectric loss in the resonator capacitor. Losses in amorphous dielectrics arise from resonant absorption from two-level system (TLS) defects with a dipole moment that couples to the electric field [7]. The loss tangent decreases with increasing microwave amplitude, and can be described by [3]:

$$\tan\delta \simeq \frac{tan\delta_0}{\sqrt{1 + (V/V_C)^{2-\Delta}}} \quad (6),$$

where $V_c$ is the critical voltage saturation of the TLS's, $\Delta$ is an experimentally determined constant, and $tan\delta_0$ is the unsaturated loss tangent. This non-trivial result for the power-dependent loss tangent allows us investigate the precision of the measurement technique.

As a result of the nonlinear dielectric loss, all the resonators except the most weakly coupled one, can be operated from the over-coupled regime ($Q_e \ll Q_i$) to the under-coupled one ($Q_e \gg Q_i$) simply by varying the input power drive. From the theory and analysis, one can show that $Q_i^{-1}(V)$ should be the same for all four resonators for moderate to weak coupling values.

In Fig. 3(b), we show data obtained in the strong coupling regime $Q_i^{-1} > 1/3 Q_e^{-1}$. In this regime, the LC and C coupled resonators exhibited $Q_i^{-1}(V)$ that deviated from that of the other two devices (one of which was weakly coupled). This indicates that the range of applicability must be considered carefully in interpreting loss data obtained on resonators with different coupling, even when external loss is accounted for.

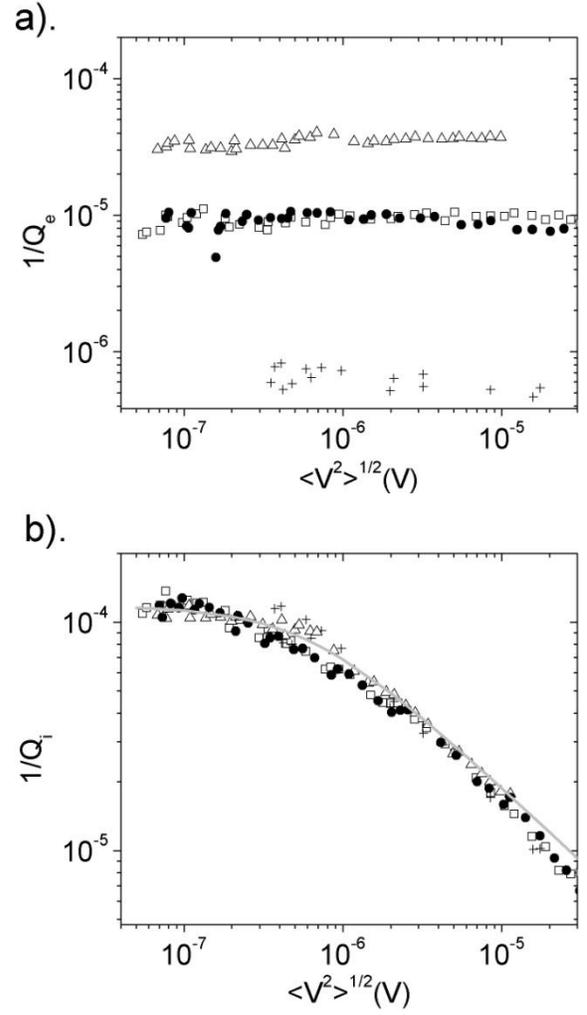

Fig. 3. Inverse internal $1/Q_i$ and external $1/Q_e$ quality factors via RMS voltage across C for resonators with different type of coupling: $\Delta$ - LC coupling; $\square$ - L coupling; $\bullet$ - C coupling; $+$ -"Weak" coupling. (a) Inverse external quality factor via V; (b) Inverse internal quality factor via V. Solid line is the TLS model calculated in accordance with Eq.(6).

We used microwave modeling software from AWR to estimate $Q_e$ and identify the resonance frequencies of each design. In the simulations, the superconducting films were approximated as perfectly conducting films. If we choose a relative dielectric constant of 6.6, we generate the values of $f_r$ and $Q_e$ shown in Table I.

TABLE I SIMULATED AND MEASURED RESONATORS PARAMETERS

| Resonator type | Experiment $f_r$, GHz | Simulation $f_r$, GHz | Experiment $Q_e$, $(10^4)$ | Simulation $Q_e$ $(10^4)$ |
|---|---|---|---|---|
| LC | 5.430 | 5.429 | 2.8 | 2.4 |
| L | 5.550 | 5.550 | 10 | 8.1 |
| C | 5.248 | 5.268 | 10 | 4.55 |
| Weak | 5.354 | 5.371 | 170 | 256 |

We found good agreement between the experimentally measured resonant frequencies and the simulations, which indicates that we can predict the resonant frequency to better than a percent in these designs, assuming we use different resonators to obtain the dielectric constant of the material used in the capacitor. From the Lorentzian shape of $S_{21}$ obtained in the simulations we also obtain an estimate for the external quality factor using:

$$Q_e = f_r/(\Delta f A) \quad (7),$$

where $\Delta f$ is resonator bandwidth, and $A$ is the resonance depth.

As Table 1 shows, we found a large difference between the LC coupled and weak coupled device, and this was consistent with the simulation. Also, the L and C coupled resonators have coupling values that are in between the values found in the strongest and weakest coupled devices. Despite this overall good agreement we find that the difference between the experimentally determined and simulated values of $Q_e$ can differ by up to a factor of 2.2, which may be partially the result of the simplifying assumptions made in the microwave simulation.

## V. Conclusion

We have measured resonators with four different types of coupling to a coplanar waveguide, We show that a minor design change provides a factor of 60 difference in the coupling between the design that was coupled with both inductance and capacitance and the design that was nominally uncoupled. Using a $SiN_x$ dielectric with power dependent loss we were able to compare results on an under coupled resonator with the three other designs which changed from over coupled to under coupled regimes. We found good agreement in all four resonator types above the boundary $Q_i^{-1} > 1/3 Q_e^{-1}$, where the coupling is from moderate to weak.